\documentclass[12pt]{iopart}
\usepackage{graphicx,epsf}
\usepackage{amssymb}
\usepackage{latexsym} 
\usepackage{amssymb}
\usepackage{color}
\begin{document}

\def\beq{\begin{equation}}
\def\eeq{\end{equation}}
\def\beqn{\begin{eqnarray}}
\def\eeqn{\end{eqnarray}}

 \title{Can we measure structures to a precision better than the Planck length?}
\author{Sabine Hossenfelder}
 
\address{NORDITA, Roslagstullsbacken 23, 106 91 Stockholm, Sweden}

\ead{hossi@nordita.org}

\begin{abstract} 

It was recently claimed that the Planck
length is not a limit to the precision by which we can measure distances, but that instead
it is merely the Planck volume that limits the precision by which we can measure volumes. 
Here, we investigate this claim and show that the argument does not support the conclusion. 

\end{abstract}
 


\section{Introduction}

Since almost 80 years now, evidence has mounted that the Planck length, $l_{\rm Pl}$, plays the r\^ole of a minimal
length or, in other words, that it sets a limit to how precisely we can measure structures. Without an experimentally
verified theory of quantum gravity, the existence of a minimal length scale is an expectation rather 
than a knowledge, yet the expectation that the bound
\beqn
\Delta x^\nu \gtrsim l_{\rm Pl} \quad, \label{usual}
\eeqn
is obeyed for spatial and temporal extensions has been supported by many thought experiments and
different approaches to quantum gravity \cite{Garay:1994en,Hossenfelder:2012jw}. 

Surprisingly, it was recently claimed by Tomassini and Viaggiu \cite{Tomassini:2011yu}, building up
on an earlier heuristic argument \cite{Doplicher:1994tu}, that the Planck
scale does not constitute a limit to the precision by which we can measure distances, but that instead
we merely have a limit to the precision by which space-time volumes can be measured
\beqn
\Delta x^0 \left( \Delta x^1 + \Delta x^2 + \Delta x^3 \right) \gtrsim l^2_{\rm Pl} \quad. \label{toma}
\eeqn
Since the inequality (\ref{toma}) follows from (\ref{usual}) 
the relevant question here is not whether (\ref{toma}) is valid, but whether (\ref{usual}) can be violated.
 
That spatial distances can be measured to a precision better than the Planck length is an extraordinary claim which, 
if correct, would mean nothing less than that arguments
dating back to Bronstein's 1936 paper \cite{Bronstein2} are all wrong. It would also pose a serious
conceptual challenge to approaches to quantum gravity which have shown indications for a minimal
length scale, such as loop quantum gravity and asymptotically safe gravity. It is the aim of this paper
to investigate Tomassini and Viaggiu's (in the following referred to as TV) argument and we will find it wanting. However, we wish to
make this a constructive criticism, and more important than pointing out why the argument is
on shaky ground, we want to clarify which steps are missing to put it on solid ground.

This paper is organized as follows. In the next section, we recall Mead's argument
 for the existence of a minimal length scale from studying the Heisenberg microscope by taking into
account general relativity. 
It is one of the most general and also most convincing arguments. In section \ref{volume} we
summarize the core of TV's argument. In section \ref{both}, we
explain the differences in the argumentation, why the case for the absence of a minimal
length does not hold up to scrutiny. We conclude in section \ref{concl}.

We use the unit convention $c=\hbar=1$, so that the Planck length $l_{\rm Pl}$ is the inverse
of the Planck mass, $m_{\rm Pl} = 1/l_{\rm Pl}$, and Newton's constant $G=l_{\rm Pl}^2$.

\section{The case for a minimal length}
\label{length}

\begin{figure}[ht]
\includegraphics[width=6.0cm]{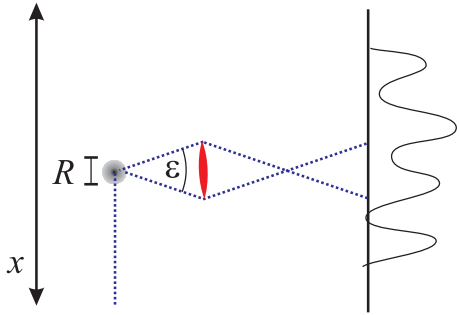}
\caption{{\small Heisenberg's microscope.}}
\label{1}
\end{figure}

Let us first recall
Heisenberg's microscope, that lead to the uncertainty principle \cite{Heisenberg3}. Consider a photon 
with frequency $\omega$ moving
in direction $x$ which scatters on a particle whose position on the $x$-axis we want to
measure. The scattered photons that reach the lens of the microscope have to lie within an
angle $\epsilon$ to produces an image from which we
want to infer the position of the particle (see figure \ref{1}). According to classical optics, the wavelength of the photon sets a limit
to the possible resolution $\Delta x$  
\beqn
\Delta x \gtrsim \frac{1}{\omega \sin \epsilon} \gtrsim \frac{1}{\omega} \quad. \label{usualdelta}
\eeqn
(Here and in the following we omit factors of order one; they do not matter for our argument.) 
But the photon used to measure the position of the particle has a recoil when it scatters 
and transfers a momentum to the particle. Since one does not know the direction of the
photon to better than $\epsilon$, this results in an uncertainty for the momentum of the
particle in direction $x$
\beqn
\Delta p_x \gtrsim \omega \sin \epsilon \quad. \label{deltap}
\eeqn 
Taken together one obtains, up to a factor of order one, Heisenberg's uncertainty \index{Uncertainty principle}
\beqn
\Delta x \Delta p_x \gtrsim 1 \quad.
\eeqn
We know today that Heisenberg's uncertainty is much more than a peculiarity of microscopy; it
is a fundamental principle of quantum mechanics. 
It does strictly speaking not even make sense to speak of
the position and momentum of the particle at the same time. Consequently, instead of speaking
about the photon scattering of the particle as if that would happen in one particular
point, we should speak of the photon having a strong interaction with the particle in
some region of size $R$.

Now we include gravity into the picture, following the treatment of Mead \cite{Mead}. 
As before, we
have a particle whose position we want to measure by help of a test particle. 
For any interaction to take place and subsequent measurement to be possible, the time
elapsed between the interaction and measurement has to be at least of the order of the
time, $\tau$, the test particle needs to travel the distance $R$, so that $\tau \gtrsim R$.
The test particle carries an energy that, though normally tiny, exerts a gravitational pull on
the particle whose position we wish to measure. It is this gravitational pull exerted
by the test particle, together with the limits by which we can know both its direction and
momentum, that causes an additional uncertainty.

The
test particle has a momentum vector $(\omega, \vec k)$, and for completeness we consider a 
particle with rest mass $\mu$, though we will see later that the tightest constraints
come from the limit $\mu \to 0$. The velocity $v$ of the test particle is
\beqn
v = \frac{k}{\sqrt{\mu^2 + k^2}}\quad, 
\eeqn
where $k^2 = \omega^2 - \mu^2$ and $k = |\vec k|$.  As before, the test particle moves into direction $x$. 
The task is now to compute the gravitational field of the test particle and the motion
it causes for the measured particle.

To obtain the metric that the test particle creates, we first change into the rest 
frame of the particle by boosting into $x$-direction. Denoting the new coordinates with
primes, the measured
particle moves towards the test particle in direction $-x'$, and the metric 
is a Schwarzschild metric. We will only need it on the
$x$-axis where we have $y=z=0$, and thus
\beqn
g'_{00} = 1 + 2 \phi' ~,~ g'_{00} = - \frac{1}{g'_{00}}~,~g'_{22}=g'_{33} = -1 ~,
\eeqn
where $\phi' = - G\mu/| x' |$, and the remaining components of the metric vanish. A Lorentz-boost back into the 
rest frame of the measured particle yields
\beqn
g_{00} &=& \frac{1 + 2 \phi}{1+ 2\phi(1-v^2)} + 2 \phi \quad,\quad g_{11} = - \frac{-1 + 2 \phi v^2}{1+2 \phi (1-v^2)} + 2 v^2 \phi \\
g_{01} &=& g_{10} = - \frac{2v\phi}{1+ \phi (1-v^2)} - 2 v \phi \quad,\quad g'_{22}=g'_{33} = -1 \quad, \label{gphi}
\eeqn
where 
\beqn
\phi = \frac{\phi'}{1-v^2} = - \frac{G\omega}{R} ~. \label{defphi}
\eeqn
Here, $R=vt -x$ is mean distance between test particle and measured particle.
To avoid a horizon in rest frame, we must have $2\phi' <1$, and thus from Eq. (\ref{defphi}) 
\beqn
- 2 \phi' = 2 \frac{G\omega}{R} (1-v^2) < 1 \quad. \label{ineqphi}
\eeqn
Because of Eq. (\ref{usualdelta}), $\Delta x \geq 1/\omega$ but also $\Delta x \geq R$ which
is the area in which the particle may scatter, and therefore 
\beqn
\Delta x^2 \gtrsim \frac{R}{\omega} \gtrsim 2 G (1 -v^2) \quad.
\eeqn
Thus, as long as $v^2 \ll 1$,  the previously found lower bound on the spatial resolution $\Delta x$ can already
be read off here, and we turn our attention towards the case where $1- v^2 \ll 1$. From (\ref{defphi}) we see that
this means we work in the limit where $-\phi \gg 1$. 

To proceed, we need to estimate now how much the measured particle moves due to the test particle's
vicinity. We denote the velocity in $x$-direction be $u$, then the requirement that the line-element ${\rm d}s^2 >0$ on
the particle's worldline yields, after some algebra, with (\ref{gphi}) the estimate
\beqn
\frac{u}{1-u} \geq - \frac{1}{2} \left(1 + 2 \phi \right) \quad. \label{uu1}
\eeqn

The time $\tau$ required for the test particle to move a distance
$R$ away from the measured particle is at least $\tau \gtrsim R/(1-u)$, and during this time the measured particle
moves a distance
\beqn
L = u \tau \gtrsim R \frac{u}{1-u} \gtrsim \frac{R}{2} \left(-1 - 2 \phi \right) \quad.
\eeqn
Since we work in the limit $- \phi \gg 1$, this means $L \gtrsim G$,
and projection on the $x$-axis yields for the uncertainty added to the measured particle
because the photon's direction was known only to precision $\epsilon$
\beqn
\Delta x \gtrsim G \omega \sin \epsilon \quad.
\eeqn
This additional uncertainty combines with (\ref{usualdelta}) 
to a lower limit for the
spatial uncertainty given by the Planck length
\beqn
\Delta x \gtrsim l_{\rm Pl}  \quad. \label{mindelta}
\eeqn

Mead continues to show that by a similar argument one finds that the precision by which
clocks can be synchronized, and thus time intervals can be measured, is also bound by the Planck
length, $\Delta x^0 \gtrsim l_{\rm Pl}$.

Adler and Santiago \cite{Adler} find the same result 
by using the linear approximation of Einstein's field equation for
a cylindrical source with length $l$ and radius $\rho$ of comparable size, filled by a radiation field with 
total energy $\omega$, and moving into direction $x$. However, this estimate can be criticized on the
grounds that the weak field approximation is strictly speaking inappropriate. 

Several other thought experiments can be found in the literature, for example 
Wigner and Salecker derived limits to the precision of time and length measurements by studying
Einstein's synchronization procedure with gravity \cite{Salecker:1957be}. Scardigli 
\cite{Scardigli:1999jh} offered a related argument from the creation and subsequent evaporation
of Planck scale black holes. Noteworthy
is also Calmet, Graesser and Hsu's argument \cite{Calmet:2004mp} for non-relativistic masses, which
has the merit of being device-independent. Limits to the measurements of black hole horizons themselves
have been studied in \cite{Wilczek,Maggiore}. They all arrive, up to a factor of order one, at the same
bounds. Ng and van Dam \cite{Ng:1994zk} argued that the scaling
behavior might be different from the one discussed here, but the lower limits remain the same. For
more details, the interested reader is referred to the recent review \cite{Hossenfelder:2012jw}. 

As before with the normal Heisenberg microscope, the relevance of (\ref{mindelta}) is not one for
microscopy. The microscope is only a placeholder for any scattering process.
In fact, at the energies needed to probe the Planck scale, the test particle almost certainly
would not scatter elastically. Instead, we should imagine a fixed-target collider experiment,
in which we try to test for a possible substructure below the Planck scale. The inequality (\ref{mindelta}) 
tells us that this is not possible. This then raises the question if not a quantum theory
that takes into account gravity should have built in this finite position uncertainty, an idea
that has received a lot of attention since Snyder \cite{Snyder} showed that it need not be in conflict
with Lorentz invariance.

\section{The case for a minimal volume}
\label{volume}

The observant reader will have noticed that the above estimate made use of 
spherical symmetry for the gravitational field of the test particle. Adler and Santiago \cite{Adler} employed cylindrical symmetry;  
however, also there it was assumed that the length and the radius of the cylinder
are of comparable size. In fact, all the other thought experiments that arrive at
the conclusion that the Planck length sets a limit to spatial resolution make implicitly
or explicitly use of spherical symmetry, in most cases by using a condition that
the extension of a mass distribution be larger than its Schwarzschild radius.

In the general case however, when the dimensions of the test particle in different 
directions are very unequal, the Hoop conjecture \index{Hoop conjecture} does not forbid any
one direction to be smaller than the Schwarzschild radius to prevent
collapse of some matter distribution, as long as at least
one other direction is larger than the Schwarzschild radius. Leaving aside that it is called a conjecture because it is unproven and just taking it
at face value, the question
then arises what limits on spatial resolution can we still
derive in the general case. 

A heuristic motivation of the following argument can be found in \cite{Doplicher:1994tu}, but
we follow here the more detailed argument by Tomassini and Viaggiu \cite{Tomassini:2011yu}.
In the absence of spherical symmetry, one may still use Penrose's isoperimetric-type conjecture \cite{Gibbons:1997js,Gibbons:2009xm}, 
according to which the area of the apparent horizon is always smaller or equal than the
event horizon, which in turn is smaller or equal than $16 \pi G^2 \omega^2$, where $\omega$
is as before the energy of the test particle. 

Now the requirement that no black hole ruins our ability to resolve short 
distances is
weakened. Instead of the requirement that the energy distribution has a 
radius larger than
the Schwarzschild radius, we only have the requirement that the area $A$, 
which encloses
$\omega,$ is large enough to prevent Penrose's condition for horizon 
formation: 
\beqn
A \geq 16 \pi G^2 \omega^2 \quad. \label{16}
\eeqn

The test particle interacts during a time $\Delta x^0$ that, by the normal uncertainty principle, 
is larger than $1/(2 \omega)$. Taking into account this uncertainty on the energy, one has
\beqn
A (\Delta x^0)^2 \geq 4 \pi G^2 \quad. \label{this2}
\eeqn
Now we have to make some assumption for the geometry of the object which will inevitably be a crude
estimate. While an exact bound will depend on the shape of the matter distribution, we will here
just be interested in obtaining a bound that depends on the three different spatial extension and
is qualitatively correct. To that end, we assume 
the mass distribution fits into some smallest box with side-lengths $\Delta x^1, \Delta x^2, \Delta x^3$, which
is similar to the limiting area
\beqn
A \sim \frac{\Delta x^1 \Delta x^2 + \Delta x^1 \Delta x^3 + \Delta x^2 \Delta x^3}{\alpha^2} \quad,
\eeqn
where we added some constant $\alpha$ to take into account different possible geometries. A comparison with the 
spherical case, $\Delta x^i = 2 R$, fixes $\alpha^2 = 3/\pi$ (we divert in the choice of this constant
from \cite{Tomassini:2011yu}, but this will not be relevant for our argument).  With Eq. (\ref{this2}) one then obtains
\beqn
\left(\Delta x^0 \right)^2 \left( \Delta x^1 \Delta x^2 + \Delta x^1 \Delta x^3 + \Delta x^2 \Delta x^3 \right) 
\geq 12 l_{\rm p}^4 \quad.
\eeqn
Since 
\beqn
\left( \Delta x^1 + \Delta x^2 + \Delta x^3 \right)^2 \geq \Delta x^1 \Delta x^2 + \Delta x^1 \Delta x^3 + \Delta x^2 \Delta x^3 
\eeqn
one also has 
\beqn
\Delta x^0 \left( \Delta x^1 + \Delta x^2 + \Delta x^3 \right) 
\geq l_{\rm p}^2 \quad.
\eeqn
Thus, as anticipated, taking into account that a black hole must not necessarily form if the spatial extension of a matter
distribution is smaller than the Schwarzschild radius into only one direction, the uncertainty we arrive at here depends on
the extension into all three directions, rather than applying separately to each. 

\section{Minimal volume or minimal length?}
\label{both}

Let us now compare the argument for a minimal length from section \ref{length} with the argument against a
minimal length from section \ref{volume}. As mentioned earlier, that there is a bound on volumes is not
the relevant statement, since this follows from the bound on the length and time intervals. Relevant
is the question whether the bound (\ref{usual}) on the length can be violated.

First we note that in section \ref{volume} one has replaced $\omega$ by the 
inverse of $\Delta x^0$, rather than combining with Eq. (\ref{usualdelta}), but that is merely a
matter of presentation and could have been done in section \ref{length} as well.

More importantly, TV's argument is not operational. The quantities
$\Delta x^\nu$ that they derive bounds on are not measurement outcomes. 
In section \ref{length}, the $\Delta x$ that we found to be limited by the Planck length
is the precision by which one can measure the position of a particle (or the presence of
substructures) with help of the test particle;
it has a clear physical interpretation. 
In section \ref{volume}, the $\Delta x^i$ are the smallest possible extensions of the 
test-particle (in the rest frame),
which with spherical symmetry would just be the Schwarzschild radius. What is, crucially, missing 
in this argument is the step in which one studies the motion of the measured particle that is 
induced by the gravitational field of the, no longer spherically symmetric, test particle.

There is another aspect of this non-operational investigation by TV. This
is the question what, fundamentally, is the test particle and how can it substantially deviate 
from spherical symmetry without making additional assumptions about the UV completion of the 
theory. We may think of a particle with a non-spherical probability distribution, one that
extends to a large distance into a direction perpendicular to the $x$-axis, so that, by TV's
argument, there is in principle nothing prohibiting us from approaching the measured particle
arbitrarily. However, if the test particle does interact with the measured particle on some
possible paths it can take, should we not expect its gravitational field to be that of a particle 
on the path rather than that of the distribution over all paths (as one would expect in
semi-classical gravity)? 

Of course this raises the question what is the gravitational field of a quantum superposition
and what happens with it upon collapse -- a question that strictly speaking is unsolved, and
will only be solved by a theory of quantum gravity. Indeed, this problem is very similar to
Hannah and Eppley's thought experiment \cite{EH} which purpose it was to show that the gravitational
field must have quantum properties like the particle which it is created by. It seems
that for TV's argument to hold, the particle's gravitational field would have to remain being
spread out even after the interaction has taken place, to avoid a strong, spherically symmetric,
gravitational field that delivers the distortion derived by Mead. This seems possible only if one
assumes that, fundamentally, the particle is not a particle but a spatially spread-out object. 
At the very least, it is not clear exactly what is being measuring and how.

Finally, let us consider a concrete example for a gravitational field of the sort that TV's 
argument would be relevant for. The gravitational field of a line mass of finite length
is the $\gamma$-metric and in the limit of infinite extension one obtains the 
Levi-Civita metric \cite{Herrera:1998eq}. In cylinder coordinates $r,z,\phi$, it takes the form \cite{Hiscock:1985uc}
\beqn
{\rm d}s^2 = -{\rm d}t^2 + {\rm d}z^2 + {\rm d}r^2 + (1-4 G \mu)^2 r^2 {\rm d} \phi^2  \quad.
\eeqn

The Levi-Civita metric is well-known for describing 
the gravitational field of a cosmic string with mass density and tension $\mu$. This
metric has no horizon. It has in fact the peculiar property of being flat. It just 
has a deficit angle of $4 \pi \mu G$: the circumference of a circle with radius $\tilde r$ is
$2 \pi \tilde r (1- 4 \mu G)$.  

What better example would there be in support of TV's argument? If we would try to
probe structures by help of such a string (at least as a limiting case)
the measured particle wouldn't even notice a gravitational field of the test string. However,
this argument would be short sighted. For even if the Levi-Civita metric has no horizon,
and we are thus not bounded by a collapse-prohibiting requirement like in the Schwarzschild
case, we still have physical considerations to take into account. If the mass density 
$\mu$ exceeds $m_{\rm Pl}^2/2$, the $r,\phi$
two-space collapses to a point \cite{Hiscock:1985uc}. 

So we know that for physical reasons the tension of the string is bounded by the square of the Planck mass. 
Now if we scatter something off the string with a momentum transfer $\omega$ that,
according to the usual uncertainty principle, was large
enough to test structures below the Planck-scale, we will transfer an energy at
least of order $\omega \sim m_{\rm Pl}$ to the string, with the direction of
momentum being perpendicular to the symmetry axis. This will cause
the string to deform to a length that we can estimate by assuming the deformation
is of triangular shape with transverse extension $\Delta x$ and base length $2 \Delta x$,
where we have assumed that the perturbation in the string travels with the same speed
that the string extends. This changes the length of the string by $(1 - \sqrt{2}) 2 \Delta x$. 
The energy in this deformation will match the transferred energy for approximately 
$\mu \Delta x = \omega$. Thus, the extension of the string in the direction
that we want to measure is
\beqn
\Delta x \sim \frac{\omega}{\mu} \gtrsim l^2_{\rm Pl} \omega  \quad.
\eeqn
Combining this with the usual uncertainty  (\ref{usualdelta}), we arrive again at Mead's conclusion 
that we cannot measure distances to better than the Planck scale.

Though the details of this argument are a little rough, this should not come as much of 
a surprise. The more energy we transfer to the string, the more it will deform, and the
higher the tension, the less it will transform. We have no other dimensionful scale at
our disposal than the Planck scale. Thus, even without knowing the details, one can
tell that the uncertainty in transverse direction will have a minimum at the Planck length.
And so, even though TVs argument is correct for what the gravitational field of the test particle
is concerned, it does not follow from this alone that one can measure structures to arbitrary
precision.

One finds a similar extension for the quantized string in string theory \cite{Susskind:1993ki,Susskind:1993aa}.
It has been argued however in \cite{Yoneya:2000bt}, that the limit on $\Delta x$ might be
avoided in string theory of one considers $D$-brane scattering, in which case $\Delta x$ could be made arbitrarily small
on the expense of making the interaction time $\Delta t$ arbitrarily large, so that merely 
a space-time uncertainty relation
\beqn
\Delta x \Delta t \sim l_{\rm s}^2 
\eeqn
is valid. (Here $l_{\rm s}$ is the string scale and in general different from the Planck scale.) 
However, at this point we have departed quite far already from a generally applicable argument and 
ventured into the realms of one particular approach to quantum gravity. 

\section{Conclusion}
\label{concl}
We have shown that the argument put forward by Tomassini and Viaggiu, according to which space-time
volumes are bounded but not spatial distances, is incomplete. But from our discussion we can now also
see what is necessary to complete the argument: An operational way to measure structures at arbitrarily
short distances that does violate the Planckian bound, presumably by use of a non-spherical geometry.
It would be interesting to see exactly which assumptions about the matter content or its quantum
properties are necessary, and to explore the physical consequences. Such an investigation might
prove insightful for understanding the r\^ole of the Planck scale in different approaches to
quantum gravity.

To answer the question posed in the title: Without the additional assumption that
extended objects exist in the fundamental description of nature, we presently do not know of any thought experiment 
that would allow to measure structures to a precision better than the Planck length.

\section*{Acknowledgements}

I thank Tomassini and Viaggiu for the thought stimulating paper and hope that my article does
appear like the constructive contribution it is meant to be, rather than a criticism. 
I thank Stefan Scherer for helpful feedback.



\begin{thebibliography}{99}

\bibitem{Garay:1994en} 
  L.~J.~Garay,
  Int.\ J.\ Mod.\ Phys.\ A {\bf 10}, 145 (1995)
  [gr-qc/9403008].

\bibitem{Hossenfelder:2012jw} 
  S.~Hossenfelder,
  arXiv:1203.6191 [gr-qc].

\bibitem{Tomassini:2011yu} 
  L.~Tomassini and S.~Viaggiu,
  Class.\ Quant.\ Grav.\  {\bf 28}, 075001 (2011)
  [arXiv:1102.0894 [gr-qc]].

\bibitem{Doplicher:1994tu}
  S.~Doplicher, K.~Fredenhagen and J.~E.~Roberts,
  Commun.\ Math.\ Phys.\  {\bf 172}, 187 (1995)
  [arXiv:hep-th/0303037].

\bibitem{Bronstein2} 
M.~Bronstein, 
 Phys. Z. Sowjetunion 9 140-157 (1936).


\bibitem{Mead} 
  C.~A.~Mead,
  Phys.\ Rev.\  {\bf 135}, B849-B862 (1964).

\bibitem{Heisenberg3} W.~Heisenberg, 
{\it ``The Physical Principles of the Quantum Theory''}, translated by C.~Eckart and F.~C.~Hoyt,  Dover, (1930).


\bibitem{Adler} 
  R.~J.~Adler, D.~I.~Santiago,
  Mod.\ Phys.\ Lett.\  {\bf A14}, 1371 (1999),  [gr-qc/9904026].

\bibitem{Salecker:1957be} 
  H.~Salecker and E.~P.~Wigner,
  Phys.\ Rev.\  {\bf 109}, 571 (1958).


\bibitem{Scardigli:1999jh}
  F.~Scardigli,
  Phys.\ Lett.\  {\bf B452}, 39-44 (1999).
  [arXiv:hep-th/9904025 [hep-th]].

\bibitem{Calmet:2004mp}
  X.~Calmet, M.~Graesser, S.~D.~H.~Hsu,
  Phys.\ Rev.\ Lett.\  {\bf 93}, 211101 (2004),
  [hep-th/0405033].

\bibitem{Wilczek} 
  S.~R.~Coleman, J.~Preskill, F.~Wilczek,
  Nucl.\ Phys.\  {\bf B378}, 175-246 (1992),
  [hep-th/9201059].

\bibitem{Maggiore} 
  M.~Maggiore,
  Phys.\ Lett.\  {\bf B304}, 65-69 (1993),
  [hep-th/9301067].


\bibitem{Ng:1994zk}
  Y.~J.~Ng and H.~van Dam,
  Annals N.\ Y.\ Acad.\ Sci.\  {\bf 755}, 579 (1995),
  [arXiv:hep-th/9406110].

\bibitem{Snyder} 
H.~S.~Snyder, 
Phys.\ Rev.\ {\bf 71}, 38 (1947).

\bibitem{Gibbons:1997js} 
  G.~W.~Gibbons,
  Class.\ Quant.\ Grav.\  {\bf 14}, 2905 (1997)
  [hep-th/9701049].

\bibitem{Gibbons:2009xm} 
  G.~W.~Gibbons,
  arXiv:0903.1580 [gr-qc].

\bibitem{EH} 
K.~Eppley and E.~ Hannah. 
Foundations of Physics, 7:51–65, (1977).


\bibitem{Herrera:1998eq} 
  L.~Herrera, F.~M.~Paiva and N.~O.~Santos,
  J.\ Math.\ Phys.\  {\bf 40}, 4064 (1999)
  [gr-qc/9810079].

\bibitem{Hiscock:1985uc} 
  W.~A.~Hiscock,
  Phys.\ Rev.\ D {\bf 31}, 3288 (1985).

\bibitem{Susskind:1993ki}
  L.~Susskind,
  Phys.\ Rev.\ Lett.\  {\bf 71} (1993) 2367
  [arXiv:hep-th/9307168].

\bibitem{Susskind:1993aa}
  L.~Susskind,
  Phys.\ Rev.\  D {\bf 49}, 6606 (1994)
  [arXiv:hep-th/9308139].

\bibitem{Yoneya:2000bt}
  T.~Yoneya,
  Prog.\ Theor.\ Phys.\  {\bf 103}, 1081-1125 (2000).
  [hep-th/0004074].

\end{thebibliography}
\end{document}